\documentclass[12pt]{article}
\usepackage{fullpage}
\usepackage{amsmath}
\usepackage{amsfonts}
\usepackage[dvips]{epsfig}

\def\##1{\underline{#1}}
\def\=#1{\underline{\underline{#1}}}

\def\+#1{\underline{\bf #1}}
\def\*#1{\underline{\underline{\bf #1}}}

\def\.{\mbox{ \tiny{$^\bullet$} }}

\def\eps{\epsilon}
\def\epso{\epsilon_{\scriptscriptstyle 0}}
\def\muo{\mu_{\scriptscriptstyle 0}}
\def\ko{k_{\scriptscriptstyle 0}}

\def\c#1{\cite{#1}}
\def\l#1{\label{#1}}
\def\r#1{(\ref{#1})}

\def\le{\left(}
\def\ri{\right)}
\def\les{\left[}
\def\ris{\right]}
\def\lec{\left\{}
\def\ric{\right\}}

\begin{document}

\begin{center}
\Large{\bf {\LARGE
Depolarization volume and correlation length in the homogenization of
 anisotropic dielectric composites
}}

\normalsize
\vspace{6mm}

Tom G. Mackay\footnote{ Tel: +44 131 650 5058; fax: +44 131 650 6553;
e--mail: T.Mackay@ed.ac.uk}

\vspace{4mm}

\noindent{ \emph{School of Mathematics,  University of
Edinburgh,\\ James Clerk Maxwell Building, King's Buildings,
Edinburgh EH9 3JZ, United Kingdom.} }

\vspace{12mm}

{\bf Abstract}\end{center}

In conventional approaches to the homogenization of
 random particulate
composites, both the distribution and size of the component phase
particles are often inadequately taken into account.
Commonly, the spatial distributions are characterized
by volume fraction alone, while
the electromagnetic response of each component particle is
represented as a vanishingly small depolarization volume.
The
strong-permittivity-fluctuation theory (SPFT) provides an alternative
approach to homogenization wherein a comprehensive description of
distributional statistics of the component phases is accommodated.
The bilocally-approximated
 SPFT is presented here for the anisotropic homogenized composite which
arises from component phases comprising ellipsoidal particles. The
distribution
of the component phases is characterized by  a two-point correlation
function and its associated correlation length. Each
 component phase particle is represented as an ellipsoidal
depolarization
region of nonzero volume.
The effects of depolarization volume and correlation length are investigated
through considering representative numerical examples.
It is demonstrated that both the spatial extent of the component phase
particles and their spatial distributions are
important factors in estimating coherent scattering losses of the macroscopic field.

\vspace{8mm}

\noindent
{\bf Keywords:} Strong-permittivity-fluctuation theory,
anisotropy, ellipsoidal particles, Bruggeman formalism

\noindent PACS numbers: 83.80.Ab, 05.40.-a


\newpage

\section{Introduction}

Consider the propagation of electromagnetic radiation through a
composite medium comprising a random distribution of particles.
Provided wavelengths are sufficiently long compared with the
 dimensions of the component  particles, the
composite may be  regarded  as an
effectively homogenous medium. The estimation of the constitutive
parameters of such a homogenized composite medium (HCM) is a
matter of long-standing scientific and technological importance
\c{L96, M03_SPIE}.
Furthermore, recent advances relating to   HCM-based
metamaterials
have  promoted interest in this area and
 highlighted the necessity for accurate  formalisms to estimate
the constitutive parameters of complex HCMs \c{Walser}.

The microstructural details of the component phases are often
inadequately
incorporated in homogenization formalisms. In particular,
conventional approaches to homogenization, as exemplified
by the widely-applied Maxwell Garnett and Bruggeman formalisms,
generally involve simplistic treatments of the distributional
statistics and sizes of the  component phase particles \c{Michel00}.
 It is noteworthy that both the Maxwell--Garnett and Bruggeman
formalisms arise in the long--wavelength regime. The
long--wavelength derivation of the Maxwell--Garnett formalism follows a
from rigorous treatment of the singularity of the free--space dyadic Green
function \c{Faxen, LW93}. The Bruggeman formalism emerges from the
 strong-permittivity-fluctuation theory (SPFT) under the
long--wavelength approximation \c{Ryzhov, TK81}.

The SPFT provides
an alternative approach to homogenization in which a comprehensive
description
of the distributional statistics of the component phases is accommodated.
Though the SPFT was originally developed for wave propagation in
continuous random mediums \c{Ryzhov, TK81, Frisch, ZL99},
 it has more recently gained prominence  in
the homogenization of particulate composites
\c{Genchev,Z94,ML95,MLW00,ML04a}.
Within the SPFT,  estimates of  the HCM constitutive parameters
are calculated as
successive refinements to the constitutive parameters of
a homogenous comparison medium. Iterates are expressed
in terms of correlation functions describing the spatial distribution
of the component phases. Correlation functions of arbitrarily high
order may be incorporated in principle, but the SPFT is usually
implemented at the second order level, known as the bilocal approximation.
A two-point correlation function and
its associated correlation length characterize the component phase
distributions for
 the second order SPFT. At lowest order (i.e.,
zeroth and first order), the SPFT estimate of HCM permittivity
is identical to that of the  Bruggeman homogenization formalism \c{M03_SPIE}.

Depolarization dyadics are key ingredients in
homogenization analyses as they characterize the electromagnetic
field inside inclusions (i.e., component phase particles)
embedded within a homogenous background.
Commonly, the inclusion particles are  treated as vanishingly
small point-like
entities; under this approximation, the corresponding  depolarization dyadic
is represented by  the singularity of the associated
dyadic Green function \c{M97, MW97}. Through neglecting the spatial
extent of the inclusions, potentially important information
is lost,  particularly when coherent scattering losses are to be
considered \c{Doyle, Dungey}.
It is noted that extended versions of both the Maxwell Garnett
formalism \c{Prinkey, SL93a}
and the Bruggeman formalism \c{Prinkey, S96} have been developed in which a nonzero
volume is attributed to the
component phase particles. However, these analyses  apply only to
isotropic HCMs and adopt a simplistic description of
 the distributional statistics of the
component phases.

In the present study, the SPFT is presented for an anisotropic
dielectric HCM.
The analysis takes place within the frequency--domain, wherein
dissipation is signalled by constitutive parameters having
nonzero
imaginary parts.
The component phases are taken to  consist of
randomly distributed, ellipsoidal particles of nonzero volume.
In Section~2,
the depolarization dyadic appropriate to an ellipsoidal
inclusion  of linear dimensions proportional to the size parameter
$\eta > 0$,
 embedded within an anisotropic comparison medium,
 is developed.
Details of the SPFT-based homogenization are presented in Section~3.
In Section~4, the theoretical results are illustrated by representative numerical
examples relating to both nondissipative and dissipative component
phases.
In particular,
nondissipative component
phases giving rise to
dissipative HCMs~---~through radiative
scattering loss from the macroscopic coherent field
\c{Kranendonk}~---~is
highlighted.
A few concluding remarks are provided in Section~5.

In the notation adopted, vector quantities are underlined whereas
3$\times$3 dyadics are double underlined.
 The inverse, adjoint, determinant and trace of
a dyadic $\=M$
are denoted by $\=M^{-1}$, $\=M^{adj}$ ,
$\mbox{det}\,\les \, \=M \, \ris $ and $\mbox{tr}\,\les \, \=M \,
\ris $, respectively.
The  identity dyadic
 is represented by  $\,\=I\,$.
All field--related quantities are implicitly
functions of the angular frquency $\omega$.
 The permittivity and
permeability of free space are denoted as $\epso$ and $\muo$,
respectively; the
free-space wavenumber is $\ko = \omega \sqrt{ \epso \muo }$.
The real and imaginary parts of  complex-valued quantities are
represented by $\mbox{Re} \,\le \. \ri$ and $\mbox{Im} \,\le \. \ri$,
respectively.

\section{Depolarization}

Let us consider an ellipsoidal inclusion of volume $V^{\eta}_e$,
orientated with its principal axes lying along the
 Cartesian $x, y, z $ axes.
 The
ellipsoidal surface  of $ V^{\eta}_e$ is parameterized
 by
\begin{equation}
\#r_{\,e}(\theta,\phi) = \eta \, \=U\.\#{\hat r} \, (\theta,
\phi),
\end{equation}
where  $\#{\hat r} \, (\theta, \phi)$ is the radial unit vector
specified by the spherical polar coordinates $\theta$ and $\phi$.
The
diagonal shape dyadic
\begin{equation}
\=U =
\frac{1}{\sqrt[3]{U_x U_y U_z}}
\;
\mbox{diag}
(U_x,
U_y, U_z), \hspace{25mm}
(U_x,
U_y, U_z > 0), \l{U}
\end{equation}
 maps the  spherical
region
 $V^\eta$ of radius $\eta$ onto the ellipsoidal region  $V^{\eta}_e$.
 The linear  dimensions of the ellipsoidal inclusion, as determined by $\eta$, are
assumed to be
sufficiently small that
the electromagnetic long-wavelength regime pertains, but not
vanishingly small.

Suppose the ellipsoidal inclusion is embedded within a homogenous
comparison medium.
In consonance with the  shape
dyadic \r{U}, the comparison medium is taken as an anisotropic
dielectric
medium characterized by  the diagonal permittivity dyadic
$\=\eps_{\,cm}$ with principal axes  aligned with those of $\=U$.
We remark that $\=\eps_{\,cm}$, being established in the
  frequency--domain, is implicitly a fuction of $\omega$.
The electromagnetic response of the ellipsoidal inclusion is  provided
by
the depolarization dyadic
\begin{equation}
\=D = \int_{V^{\eta}_e} \, \=G_{\,cm} (\#r) \; d^3 \#r \,= \int_{V^\eta} \, \=G_{\,cm}
(\=U \. \#r) \; d^3 \#r \,. \l{depol_def}
\end{equation}
Herein,  $\=G_{\,cm} (\#r)$ is the  dyadic Green function of the comparison medium which
satisfies the nonhomogenous vector Helmholtz equation
\begin{equation}
\le \nabla \times \nabla \times \=I
 - \omega^2 \muo \, \=\eps_{\,cm} \ri \. \=G_{\,cm} (\#r -
\#r') =  i \omega \muo \delta \le \#r - \#r' \ri \=I\,. \l{Helm}
\end{equation}
Explicit representations of  Green functions are  not generally
available for anisotropic mediums
\c{W93}. However, it suffices for our present purposes to consider
the Fourier transform of  $\=G_{\,cm} (\#r)$, namely
\begin{equation}
\={\tilde{G}}_{\,cm} (\#q) = \int_{\#r} \=G_{\,cm} (\#r) \, \exp (- i \#q \. \#r )
\; d^3 \#r \,.
\end{equation}
Taking the Fourier transform of equation \r{Helm} delivers
$\={\tilde{G}}_{\,cm} (\#q)$ as
\begin{equation}
\={\tilde{G}}_{\,cm} (\#q) = - i \omega \muo \le
\#q \times \#q \times \=I  + \omega^2 \muo \, \=\eps_{\,cm} \ri^{-1}.
\end{equation}
Thereby, equation \r{depol_def} yields
 \c{M97, MW97}
\begin{eqnarray}
\=D &=&
 \frac{\eta}{2 \pi^2} \, \int_{\#q} \frac{1}{q^2} \, \le \frac{\sin
(q \eta )}{q \eta} - \cos ( q \eta) \ri \, \={\tilde{G}}_{\,cm} (\=U^{-1}\.\#q)
\; d^3 \#q \, .
\end{eqnarray}
In order to  consider the depolarization of an inclusion of
nonzero volume, we  express  $\=D$ as the sum
\begin{equation}
\=D = \=D^\eta + \=D^{ 0} \,
\end{equation}
where
\begin{eqnarray}
&&
\=D^\eta =  \frac{\eta}{2 \pi^2} \, \int_{\#q} \frac{1}{q^2} \, \le \frac{\sin
(q \eta )}{q \eta} - \cos ( q \eta) \ri \, \={\tilde{G}}^{\eta}_{\,cm} (\=U^{-1}\.\#q)
\; d^3 \#q \, , \l{D0} \\
&&
\=D^{ 0} =  \frac{\eta}{2 \pi^2} \, \int_{\#q} \frac{1}{q^2} \, \le \frac{\sin
(q \eta )}{q \eta} - \cos ( q \eta) \ri \, \={\tilde{G}}^{ \infty}_{\,cm} (\=U^{-1}\.\hat{\#q})
\; d^3 \#q \, , \l{Dinf}
\end{eqnarray}
with
\begin{eqnarray}
\underline{\underline{\tilde{\bf G}}}^{\eta}_{\,cm}(\=U^{-1}\.\#q) &=&
\underline{\underline{\tilde{\bf G}}}_{\,cm} (\=U^{-1}\.\#q) -
\underline{\underline{\tilde{\bf G}}}^{\infty}_{\,cm} (\=U^{-1}\.\hat{\#q})\,, \l{G_Go_Gi}
\\
\underline{\underline{\tilde{\bf G}}}^{\infty}_{\,cm}(\=U^{-1}\.\hat{\#q})
&=& \lim_{q\rightarrow \infty} \;  \underline{\underline{\tilde{\bf G}}}_{\,cm} (\=U^{-1}\.\#q).
\end{eqnarray}
Thus, the dyadic  $ \=D^{0}$ represents the depolarization contribution
arising from
the vanishingly small region
$ \displaystyle{\lim_{\eta \rightarrow 0}} V^{\eta}_e $,
whereas  the dyadic $ \=D^{\eta}$ provides the depolarization contribution
arising from the region of nonzero volume $ \le V^{\eta}_e -
 \displaystyle{\lim_{\eta \rightarrow 0}} V^{\eta}_e \ri
 $.
In homogenization studies, it is common practice to
neglect
 $ \=D^{\eta}$ and
assume that the
depolarization dyadic is given by  $ \=D^{0}$ alone \c{M03_SPIE}.
However, studies of  isotropic HCMs  have emphasized the importance of
 the nonzero spatial extent of depolarization
regions \c{Doyle, Dungey, Prinkey, SL93a, S96}.

The properties of depolarization dyadics associated with  vanishingly small
inclusions have  been widely investigated:
the volume integral \r{Dinf} simplifies to the $\eta$--independent surface integral
  \c{M97, MW97}
\begin{eqnarray}
\=D^{0} &=& \frac{1}{  4 \pi i \omega} \, \int^{2\pi}_0 \; d\phi
\, \int^\pi_0\;  d\theta\; \sin \theta \, \le \,
\frac{1}{\mbox{tr} \le \,\=\eps_{\,cm}\.\=A \,\ri} \, \=A \,
\ri\,, \l{depol}
\end{eqnarray}
 where
\begin{equation}
\=A = \mbox{diag} \,\le \, \frac{\sin^2 \theta \,
\cos^2\phi}{U^2_x},\, \frac{\sin^2 \theta \,
\sin^2\phi}{U^2_y},\,\frac{\cos^2\theta}{U^2_z}\,\ri\,.
\end{equation}
Furthermore, the integrations of \r{depol}
 reduce to elliptic function representations \c{W98}. In the
case of spheroidal inclusion  geometries, hyperbolic
 functions
provide an evaluation of $\=D^{0}$ \c{M97}, while for the degenerate
isotropic case  $U_x = U_y = U_z$  the well-known
result $\=D^0  = \le 1 /3 i \omega  \ri \=\eps^{-1}_{\,cm}$ emerges \c{BH}.

We  focus attention on $\=D^\eta$.
By analogy with an equivalent
integration which  arises within the strong--permittivity--fluctuation
theory \c{M03},  the application
  of residue calculus to \r{D0} provides
\begin{eqnarray}
&& \=D^\eta = \frac{1}{4 \pi i \omega} \, \=W(\eta), \l{D0_W}
\end{eqnarray}
where the dyadic function $\=W(\eta)$ has the surface integral representation
\begin{eqnarray}
 && \=W(\eta) = \eta^3
\int^{2 \pi}_{0}  d \phi  \int^{\pi}_0
 d \theta\;\;
\frac{\sin \theta}{3 \, \Delta}
\lec \,  \les \,
 \frac{3 \le \kappa_+ -
\kappa_-  \ri}{2 \eta} + i \le \kappa^{\frac{3}{2}}_+  - \kappa^{\frac{3}{2}}_-  \ri
\ris
  \=\alpha  +
i  \omega^2 \muo \,\le \kappa^{\frac{1}{2}}_+ -  \kappa^{\frac{1}{2}}_-  \ri  \=\beta \,
\ric, \nonumber \\ &&  \l{W_def}
\end{eqnarray}
with
\begin{eqnarray}
&&  \=\alpha =
  \les 2 \,\=\eps_{\,cm} - \mbox{tr} \le \, \=\eps_{\,cm} \, \ri
\, \=I \, \ris\. \=A - \mbox{tr} \le \,\=\eps_{\,cm}\.\=A\,\ri \,
\=I\,
 -  \, \frac{  \mbox{tr} \le \, \=\eps^{adj}_{\,cm}\.\=A\,\ri -
\les \, \mbox{tr} \le \, \=\eps^{adj}_{\,cm} \, \ri \, \mbox{tr} \le
\, \=A \, \ri \, \ris }{
 \mbox{tr} \le \, \=\eps_{\,cm}\. \=A \, \ri } \,  \=A \, \nonumber \\
&&
\\
&& \=\beta =
 \=\eps^{adj}_{\,cm} -  \frac{  \det \le \, \=\eps_{\,cm} \, \ri }
{ \mbox{tr} \le \, \=\eps_{\,cm}\. \=A \, \ri} \, \=A \,,
\\
&&  \Delta =   \lec \les \mbox{tr} \le \, \=\eps^{adj}_{\,cm}\.\=A\,\ri -
 \mbox{tr} \le \, \=\eps^{adj}_{\,cm} \, \ri \, \mbox{tr} \le
\, \=A \, \ri \, \ris^2 - 4 \det \le \, \=\eps_{\,cm} \, \ri
\mbox{tr} \le \, \=A \, \ri \,
 \mbox{tr} \le \, \=\eps_{\,cm}\. \=A \, \ri \ric^{\frac{1}{2}}, \\
&& \kappa_{\pm}  =
\muo \omega^2 \frac{
\les \, \mbox{tr} \le \, \=\eps^{adj}_{\,cm} \, \ri \, \mbox{tr} \le
\, \=A \, \ri \, \ris - \mbox{tr} \le \, \=\eps^{adj}_{\,cm}\.\=A\,\ri
\pm \Delta}{2 \, \mbox{tr} \le \, \=A \, \ri \,
 \mbox{tr} \le \, \=\eps_{\,cm}\. \=A \, \ri}.
\end{eqnarray}
The surface integrals of \r{depol} and \r{W_def} are straightforwardly
evaluated by standard numerical techniques \c{Fortran}.

\section{Homogenization }

The SPFT may be applied to estimate the constitutive parameters of
HCMs. Let us consider the homogenization of a two--phase composite
 wherein the two component phases, labelled as $a$ and $b$,
comprise  ellipsoidal particles of shape specified by
$\=U$ and linear dimensions specified by $\eta > 0$. A random
distribution of identically orientated particles is
envisaged. The component phases $a$ and $b$ are taken to be
isotropic dielectric mediums with permittivities $\eps_a$ and
$\eps_b$, respectively.

The distributional statistics of the component phases are described in
terms of moments of the characteristic functions
\begin{equation}
\Phi_{ \ell}(\#r) = \left\{ \begin{array}{ll} 1, & \qquad \#r \in
V_{\, \ell},\\ & \qquad \qquad \qquad \qquad \qquad \qquad (\ell=a,b) . \\
 0, & \qquad \#r \not\in V_{\, \ell}, \end{array} \right.
\end{equation}
 The volume fraction of phase $\ell$, namely $f_\ell$ , is given by
the first statistical moment of
 $\Phi_{\ell}$ ;
 i.e., $\langle \, \Phi_{\ell}(\#r) \, \rangle = f_\ell$ .
 Clearly,
 $f_a + f_b = 1$.
The second statistical moment of $\Phi_{\ell}$
 provides a two-point covariance function.
We adopt the physically-motivated form \c{TKN82}
\begin{equation}
\langle \, \Phi_\ell (\#r) \, \Phi_\ell (\#r')\,\rangle =
\left\{
\begin{array}{lll}
\langle \, \Phi_\ell (\#r) \, \rangle \langle \Phi_\ell
(\#r')\,\rangle\,, & & \hspace{10mm}  | \, \=U^{-1}\. \le   \#r - \#r' \ri | > L \,,\\ && \hspace{25mm} \\
\langle \, \Phi_\ell (\#r) \, \rangle \,, && \hspace{10mm}
 | \, \=U^{-1} \. \le  \#r -
\#r' \ri | \leq L\,,
\end{array}
\right.
 \l{cov}
\end{equation}
where $L>0$ is the correlation length. It is remarked that the
specific form of the covariance function has little influence on
SPFT estimates of HCM constititutive parameters  \c{MLW01b}.


The $n$th  order SPFT estimate of the HCM permittivity, namely
$
\=\eps^{[n]}_{\,ab}$,
 is based upon the
iterative refinement of the comparison medium permittivity, namely
$\=\eps_{\,cm}$. To zeroth order and first order, the SPFT
permittivity estimate is identical to the comparison medium
permittivity \c{M03_SPIE}; i.e.,
\begin{equation}
\=\eps^{[0]}_{\,ab} = \=\eps^{[1]}_{\,ab}
 = \=\eps_{\,cm}.
\end{equation}
The well--known Bruggeman homogenization formalism provides the
estimate of $\=\eps_{\,cm}$ \c{MLW00}.
 Thereby, $\=\eps_{\,cm}$ emerges through
solving the nonlinear equations
\begin{equation}
f_a \, \=\chi_{\,a} + f_b \, \=\chi_{\,b} = \=0\,,
\end{equation}
wherein the
polarization dyadics
\begin{equation}
\=\chi_{\,\ell} = -i\,\omega\,\le\,\eps_{\ell}\,\=I -
\=\eps_{\,cm}\,\ri\.\=\Gamma^{-1}_{\,\ell}\,, \hspace{30mm} (\ell
=a,b), \l{X_def}
 \end{equation}
are dependent upon the inclusion size parameter $\eta$ via
\begin{equation}
\=\Gamma_{\,\ell} =  \les\, \=I + i \omega \, \=D\.\le\,
\eps_{\ell}\,\=I - \=\eps_{\,cm}\,\ri\,\ris\,. \l{e16}
\end{equation}
Recursive procedures generate the $p$th iterate of $\=\eps_{\,cm}$ as
\begin{equation}
\=\eps_{\,cm}[p] =  \mathcal{T}
 \, \lec \,  \=\eps_{\,cm} \,[p-1] \, \ric,
\end{equation}
 with
the operator $\mathcal{T}$ being defined by
\begin{eqnarray}
\mathcal{T} \, \lec \, \=\eps_{\,cm} \,  \ric
&=& \le \, f_a \, \eps_{a}
\, \=\Gamma^{-1}_{\,a} +  f_b \,  \eps_{b} \, \=\Gamma^{-1}_{\,b} \, \ri
\.
 \le \, f_a \,  \=\Gamma^{-1}_{\,a} +  f_b \, \=\Gamma^{-1}_{\,b} \, \ri^{-1}\,.
\l{tau}
\end{eqnarray}


The SPFT is most widely implemented at the second order
level~---~known as  the bilocal approximation~---~which provides
the following estimate of the HCM permittivity dyadic \c{MLW00}
\begin{equation}
\=\eps^{[2]}_{\,ab}  =
 \=\eps_{\,cm} - \frac{1}{i \,\omega}\,\le\,\=I +
\=\Sigma \. \=D \,\ri^{-1}\.\=\Sigma\,.
\end{equation}
Thus, the inclusion size parameter $\eta$ influences
$\=\eps^{[2]}_{\,ab}$ directly through the depolarization dyadic $\=D$
and
indirectly through
 the \emph{mass operator} \c{Frisch} dyadic term
\begin{equation}
\=\Sigma =
\frac{f_a f_b }{4 \pi i \omega}
\le\,\=\chi_{\,a} -
\=\chi_{\,b}\,\ri\.\=W(L)\.\le\,\=\chi_{\,a} - \=\chi_{\,b}\,\ri.
\end{equation}
Notice that the correlation length $L$~---~which plays a key role in the
second order SPFT~---~is not relevant to the zeroth order SPFT.

\section{Numerical studies}

We investigate the  theoretical results presented in Sections
2 and 3 by means of representative numerical studies. To
highlight the particular effects of depolarization volume  and
correlation length, it is expedient to begin with a study of HCMs
arising from  nondissipative component mediums (i.e., $\eps_{a,b}
\in \mathbb{R}$) before proceeding to consider HCMs arising from
dissipative component mediums (i.e., $\eps_{a,b} \in
\mathbb{C}$ with $\mbox{Im} \, \le \eps_{a,b} \ri > 0 $).
However, we note that caution should be exercised when applying
the SPFT and Bruggeman formalism to weakly dissipative HCMs where
$\mbox{Re} \, \le \eps_a \ri $ and $ \mbox{Re} \, \le \eps_b \ri $
have opposite signs
 \c{ML04b}.
 For all calculations reported here, the angular
frequency $\omega$ is fixed at $ 2 \pi \times 10^{10}\; \mbox{rad
s}^{-1}$.

\subsection{Nondissipative component phases}

Consider, for example,  the HCM which arises
from the nondissipative component phases  specified by the
permittivities and shape parameters
\begin{equation}
\left.
\begin{array}{lll}
 \eps_a = 2 \epso, & \eps_b = 12 \epso & \\
U_x = 1, & U_y = 3, & U_z =
15
\end{array}
\right\}. \l{nondiss}
\end{equation}

In Figure~1, the components of
the zeroth order SPFT estimate
of HCM permittivity $\=\eps^{[0]}_{\,ab} = \epso \, \mbox{diag}
\le \eps^x_{ab0}, \,\eps^y_{ab0}, \,\eps^z_{ab0} \ri $ are plotted
against volume fraction $f_a$ and relative inclusion size $\ko
\eta$. The permittivity parameters  $\eps^{x,y,z}_{ab0}$
 are constrained to agree with $\eps_b$ and
$\eps_a$ at $f_a = 0$ and $f_a = 1$, respectively. For volume
fractions $f_a \in (0,1)$,
the ellipsoidal particulate  geometry of the component
phases endows the HCM with biaxial anisotropy and we observe that
 $\eps^x_{ab0}$, $\eps^y_{ab0}$, $\eps^z_{ab0}$, $\eps_a$ and
 $\eps_b$ each  take different values.
At $\eta = 0$  the estimates of the HCM permittivity parameters
 $\eps^{x,y,z}_{ab0}$
are the same as  those of the conventional Bruggeman
homogenization formalism.
When $\eta > 0$  the
 parameters $\eps^{x,y,z}_{ab0}$ become complex-valued with positive imaginary
parts,
despite
 the component phase permittivities $\eps_a$ and $\eps_b$
being  real-valued.  The manifestation of
$\mbox{Im} \, \le \eps^{x,y,z}_{ab0} \ri  > 0$
 indicates radiative scattering
loss from the macroscopic
coherent field \c{Kranendonk}. Thus, we see that by attributing a
nonzero volume to  the component phase
inclusions, coherent scattering is accommodated, thereby
giving rise to  a dissipative HCM. Since  the
imaginary parts of $\eps^{x,y,z}_{ab0}$
 are  observed to increase with increasing $\eta$, it is
deduced  that
  more radiation is scattered out of the
coherent field by  enlarging the depolarization volumes.
The real parts of the HCM permittivity parameters
 $\eps^{x,y,z}_{ab0}$ are found to be relatively insensitive to the
inclusion size parameter $\eta$.

Let us turn now to the bilocally-approximated SPFT. For
vanishingly small component phase inclusions (i.e., $\eta = 0$),
the components of the second order SPFT estimate of HCM
permittivity $\=\eps^{[2]}_{\,ab} = \epso \, \mbox{diag} \le
\eps^x_{ab2}, \,\eps^y_{ab2}, \,\eps^z_{ab2} \ri $ are graphed
against volume fraction $f_a$ and relative correlation length $\ko
L$ in Figure~2. As $L \rightarrow 0$, the second order SPFT
estimates converge to those of the zeroth order SPFT. Accordingly,
the   estimates of the HCM permittivity parameters
$\eps^{x,y,z}_{ab2}$ at $L=0$ are the same as the estimates
deriving from  the conventional Bruggeman homogenization
formalism. As the correlation length $L$ increases from 0, the HCM
parameters $\eps^{x,y,z}_{ab2}$ acquire  positive-valued imaginary
parts. Furthermore, as  $L$ increases so the imaginary parts of
the HCM parameters $\eps^{x,y,z}_{ab2}$ become increasingly large.
Thus, we infer that the scattering interactions of increasing
numbers of component phase inclusions become correlated as $L$
increases, resulting in an increasingly  dissipative  HCM. The
real parts of the  HCM permittivity parameters
$\eps^{x,y,z}_{ab2}$ do not vary significantly as the correlation
length increases, as is illustrated in Figure~2. Notice that the
relationship between the correlation length $L$ and the second
order SPFT parameters $\eps^{x,y,z}_{ab2}$ for $\eta = 0$ closely
resembles the relationship between the inclusion size parameter
$\eta$ and the zeroth order SPFT
 parameters $\eps^{x,y,z}_{ab0}$ presented in Figure~1.

The result of combining an   inclusion size parameter
$\eta > 0$  with a  correlation length $L > 0$ is presented in
Figure~3. Therein,  the components of
the second order SPFT estimate
$\=\eps^{[2]}_{\,ab}$
are plotted against both relative correlation length $\ko L$ and the
relative inclusion size $\eta / L$ for the fixed volume fraction $f_a
= 0.3$. The  real and imaginary parts of
  the HCM permittivity
parameters  $\eps^{x,y,z}_{ab2}$
are seen to increase with both $\eta$ and $L$.
Furthermore, the rate of increase of  $\mbox{Im}\, \le
\eps^{x,y,z}_{ab2} \ri$
  exceeds the corresponding rate of increase
observed in Figure~1 when $L = 0$,  and in Figure~2 when $\eta =
0$. Hence, by taking account of both  nonzero depolarization
volume and correlation length, more radiation is scattered out of
the coherent field than would be the case if either nonzero
depolarization volume  or correlation length alone was taken into
account.

\subsection{Dissipative component phases}

While it is instructive to consider nondissipative component phases  in
order to highlight the particular effects of  depolarization volume and
correlation length, in reality all mediums exhibit dissipation.
Therefore,  let us  now consider  the homogenization of
component phases characterized by the
complex-valued permittivities
$\eps_{a } = \le 2 + 0.05i \ri \epso$ and  $\eps_b
 = \le 12 + 0.4i \ri \epso$. We continue with the
same values for the inclusion shape parameters
 $U_x$, $U_y$ and $U_z$ as given in \r{nondiss}.
The volume fraction is fixed at $f_a = 0.3$ for all
results presented in this section.

In Figure~4, the components of the zeroth order SPFT estimate of
HCM permittivity $\=\eps^{[0]}_{\,ab}  $ are plotted against
relative inclusion size $\ko \eta$. Since the component phases are
themselves dissipative,  the imaginary parts of
 $ \eps^{x,y,z}_{ab0}$
are positive-valued at $\eta = 0$. As the inclusion size parameter
$\eta$
increases, so the imaginary parts of  $ \eps^{x,y,z}_{ab0}$
 also increase. The growth in $\mbox{Im}\, \le
\eps^{x,y,z}_{ab0} \ri $ is attributable to the increasing amount of
coherent scattering loss which develops as the depolarization regions
increase in size. As was observed in the corresponding instance
 for nondissipative
component phases, the real parts of $\eps^{x,y,z}_{ab0}$ are
relatively insensitive to $\eta$.

The second order SPFT estimates of the HCM permittivity parameters
$\eps^{x,y,z}_{ab2}$, calculated with $\eta = 0$, are graphed
against relative correlation length $\ko L$ in Figure~5.
The observed increase in $\mbox{Im}\, \le
\eps^{x,y,z}_{ab0} \ri $ as $L$ increases reflects the
fact that coherent losses increase as the actions of increasing
numbers of scattering centres become correlated.
The real parts of the HCM permittivity parameters vary little as the
correlation length increases.
Comparing Figures~4 and 5, it is clear that
the dependency upon $L$ for the second-order SPFT with $\eta = 0$
closely resembles the dependency
upon $\eta$ for the zeroth order SPFT.

Finally, let us consider the combined effect of
 nonzero depolarization volume   and
correlation length.  In Figure~6,
$\eps^{x,y,z}_{ab2}$
are plotted against relative inclusion size $\ko \eta$ for the fixed
relative correlation length $\ko L = 0.1$. While $\mbox{Re} \, \le
\eps^{x,y,z}_{ab2} \ri $
 are relatively insensitive to increasing
$\eta$,  $\mbox{Im} \, \le \eps^{x,y,z}_{ab2} \ri $
increase markedly as the inclusion size parameter increases.
Furthermore, the rate of increase of  $\mbox{Im} \, \le
\eps^{x,y,z}_{ab2} \ri $ in Figure~6
is greater than that seen in Figure~4 where the influence of $\eta $
alone is considered, and in Figure~5 where the influence of $L$ alone
is considered.

\section{Concluding remarks}

In  implementations of homogenization formalisms, such as the
widely-used Maxwell Garnett and Bruggeman formalisms, the
distributional statistics and sizes  of the component phase
particles are often inadequately taken into account.
 A notable
exception is provided by the SPFT in which the distributional
statistics of the component phases are described through a
hierarchy of spatial correlation functions. In the present study
the SPFT is extended through attributing a nonzero volume to the
component phase inclusions. Thereby, we show that both
depolarization volume and correlation length contribute to
coherent scattering losses. Furthermore, the effects of
depolarization volume and correlation length upon the imaginary
parts of the HCM constitutive parameters are  cumulative and
reflect the anisotropy of the HCM.
We remark that our numerical results which relate to the effect of
depolarization volume alone, as presented in
 presented in Figures~1 and 4, are consistent with
numerical results based on the extended Maxwell Garnett and
extended Bruggeman homogenization formalisms for isotropic dielectric
HCMs \c{Prinkey, SL93b}.
The importance of incorporating
microstructural details within homogenization formalisms is thus
further emphasized.

\vspace{10mm}

\noindent
{\bf Acknowledgement:} The author thanks  an anonymous referee for
suggesting improvements to the manuscript and drawing his attention to
Ref. \c{Kranendonk}.

\newpage

\begin{figure}[!ht]
\centering \psfull \epsfig{file=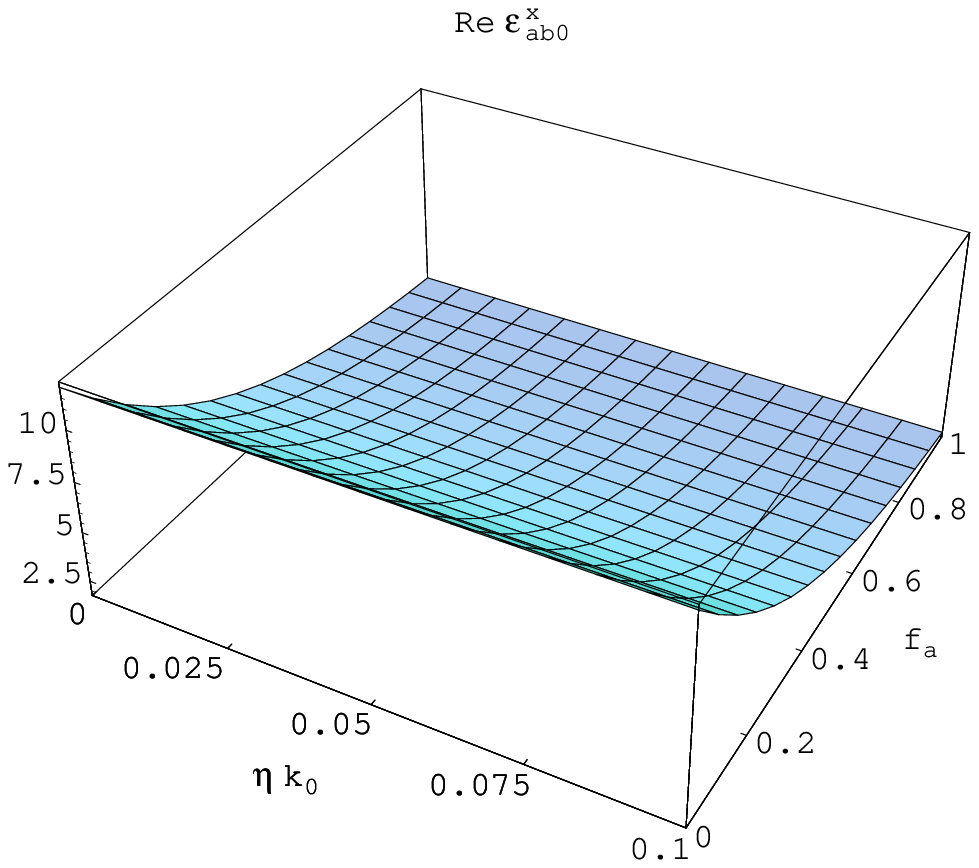,width=3.0in}
\hfill
  \epsfig{file=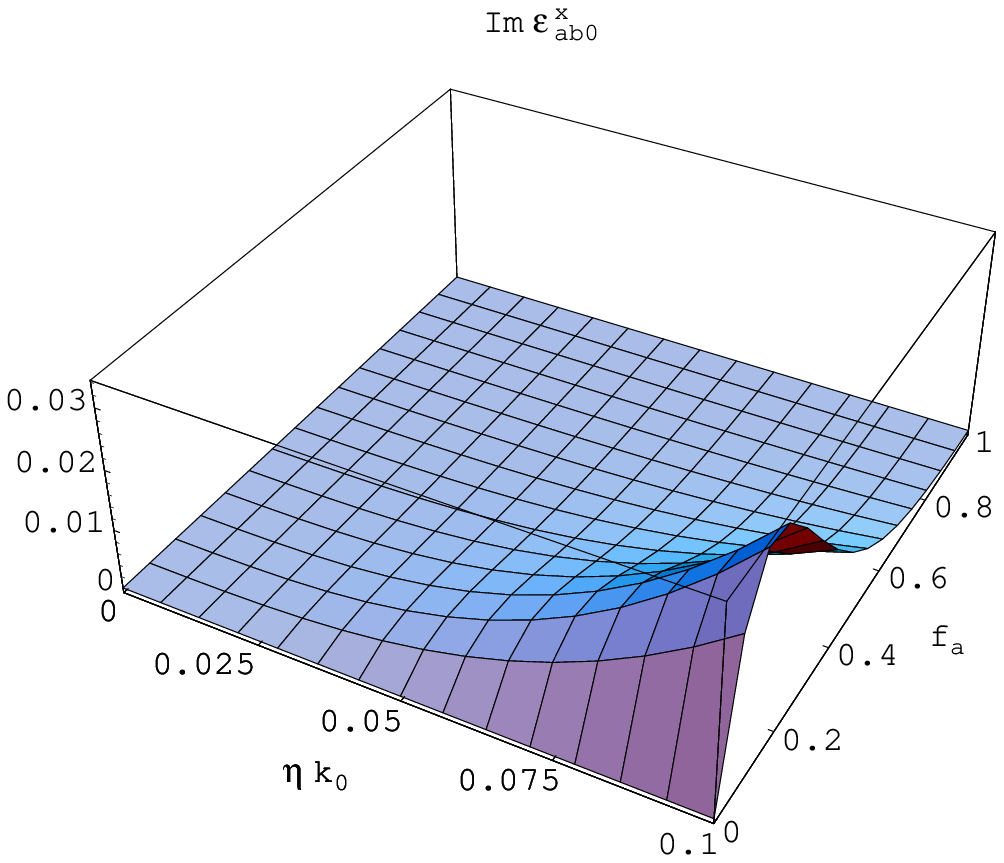,width=3.0in}
\epsfig{file=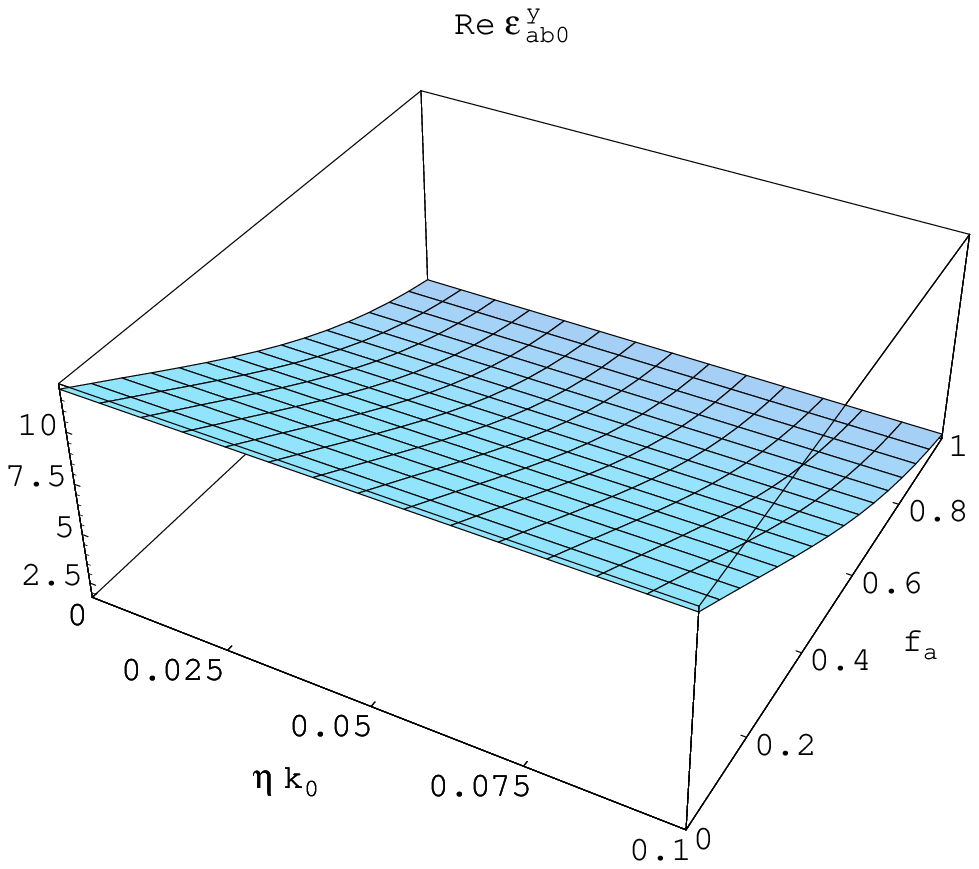,width=3.0in} \hfill
\epsfig{file=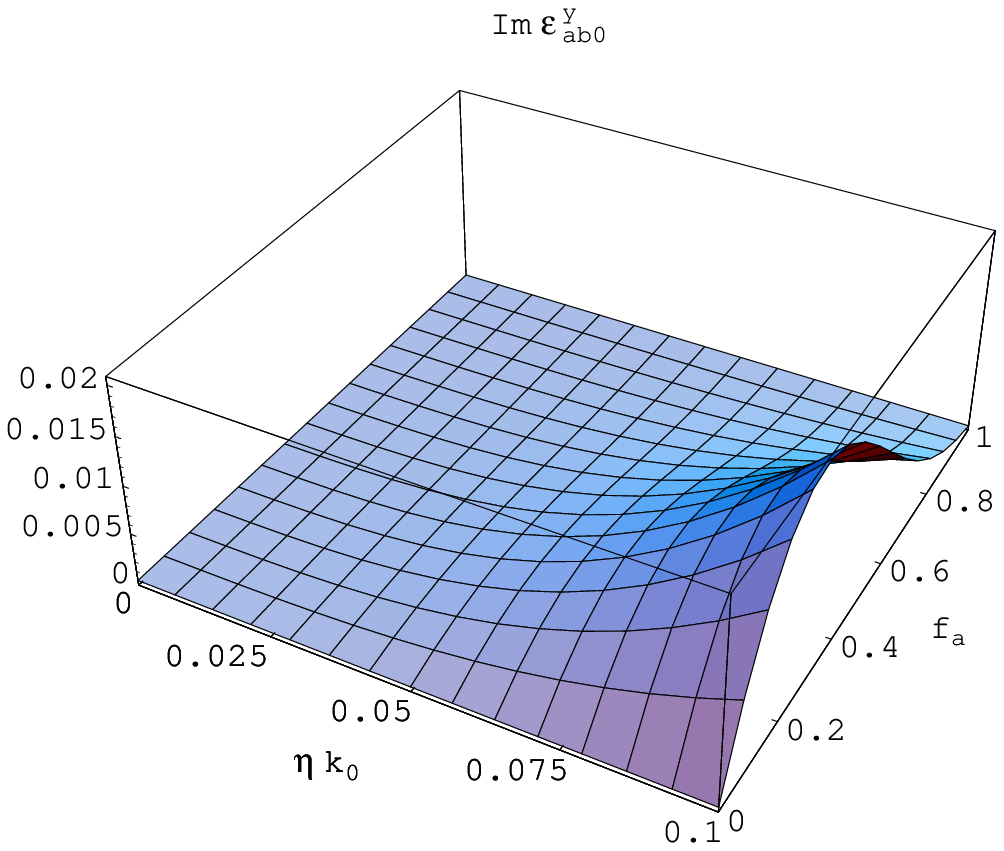,width=3.0in}
\epsfig{file=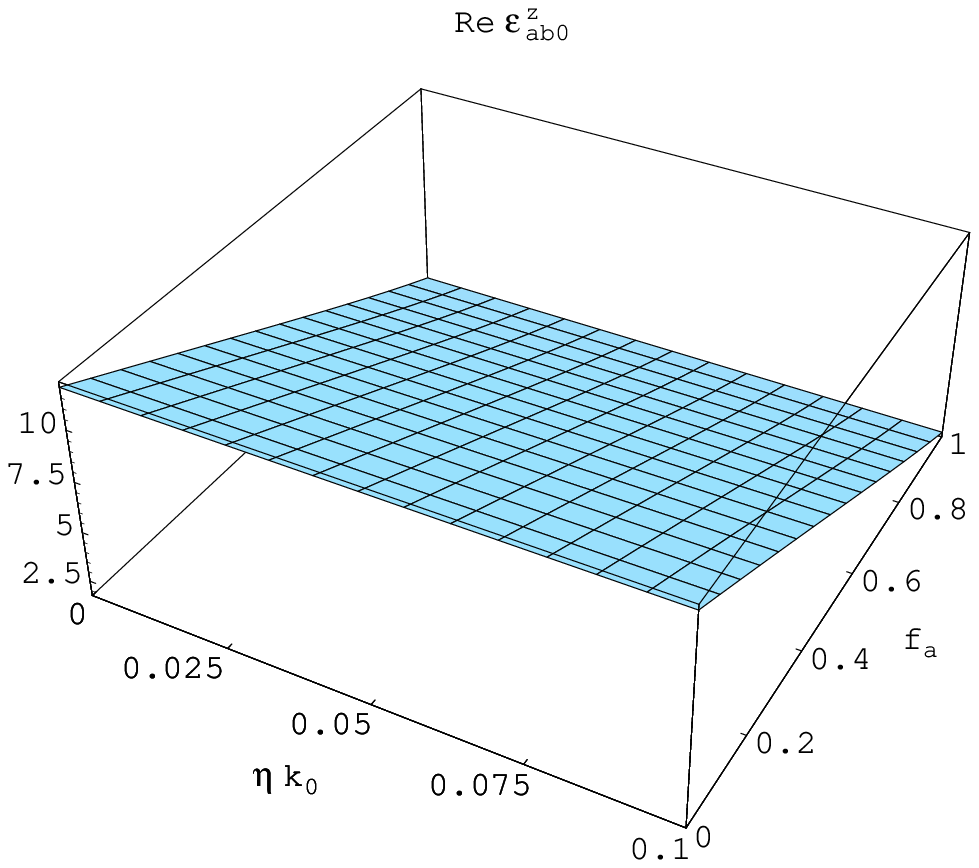,width=3.0in} \hfill
\epsfig{file=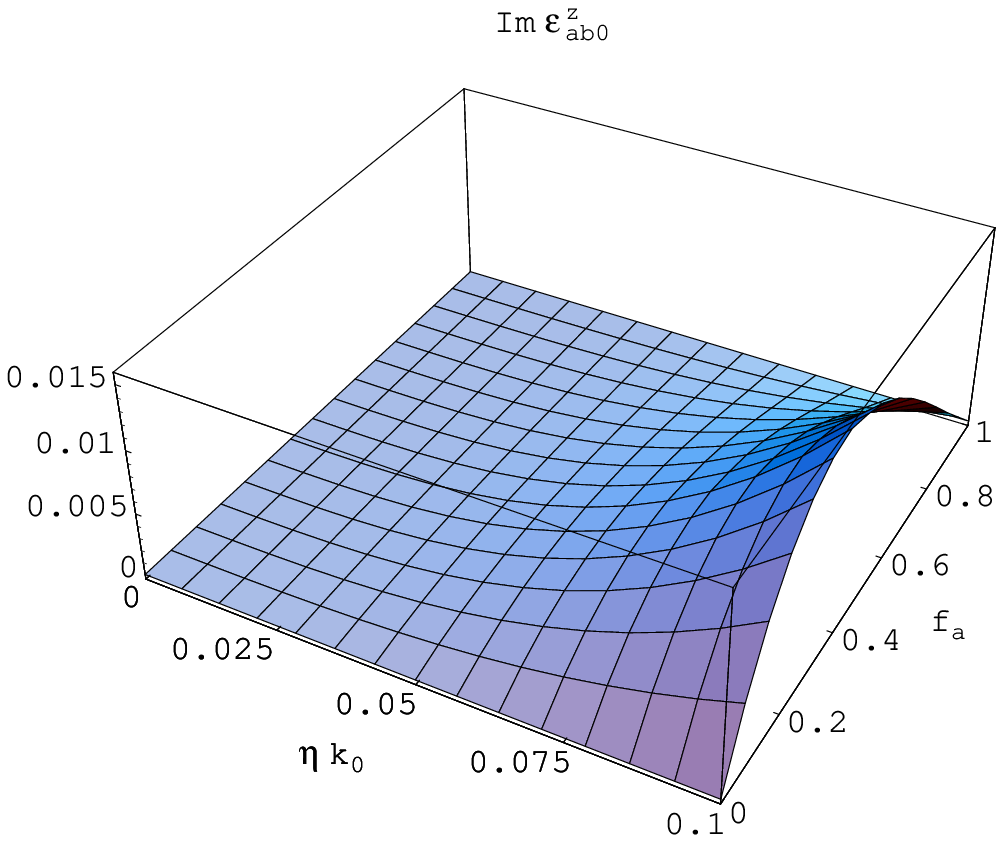,width=3.0in} \caption{ Real (left) and
imaginary (right) parts of the zeroth order SPFT estimate of HCM
permittivity $\, \underline{\underline{ \eps}}^{[0]}_{\,ab} =
\epso \, \mbox{diag} \le \eps^x_{ab0}, \,\eps^y_{ab0},
\,\eps^z_{ab0} \ri $ plotted against volume fraction $f_a$ and
relative inclusion size $\ko \eta $.
 Component phase
parameter values: $\eps_{a } = 2 \epso$,  $\eps_b = 12 \epso$,
 $U_x = 1$, $U_y = 3$ and $U_z = 15$.}

\end{figure}

\newpage

\begin{figure}[!ht]
\centering \psfull \epsfig{file=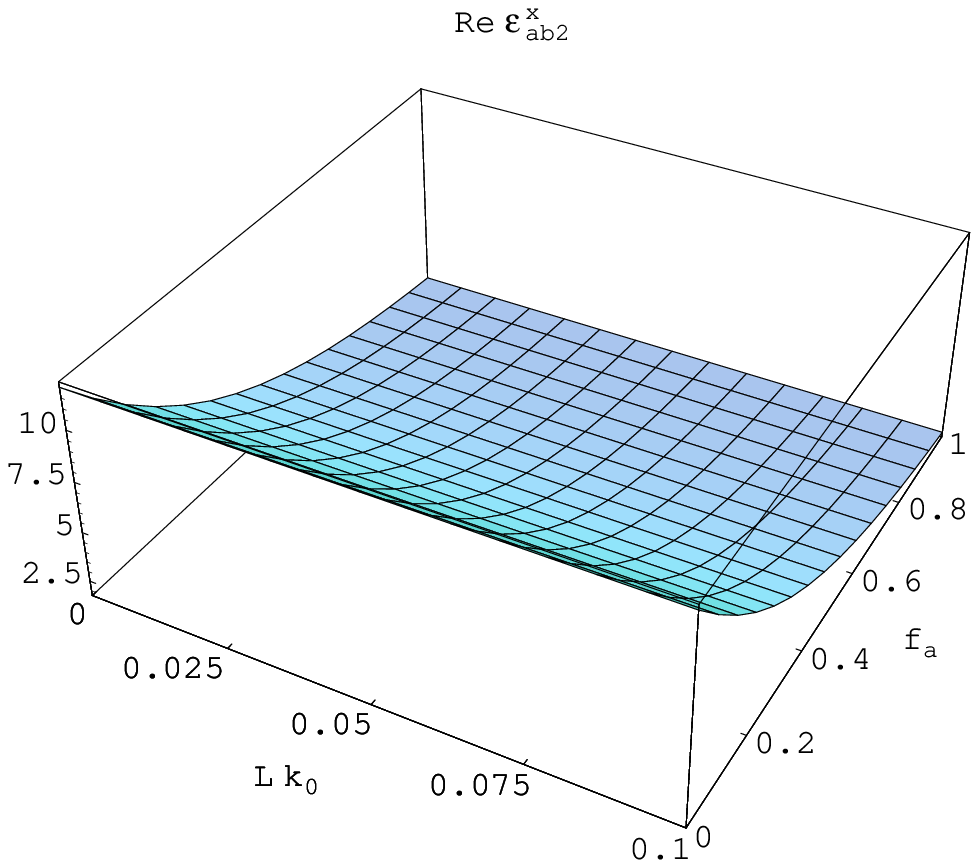,width=3.0in}
\hfill
  \epsfig{file=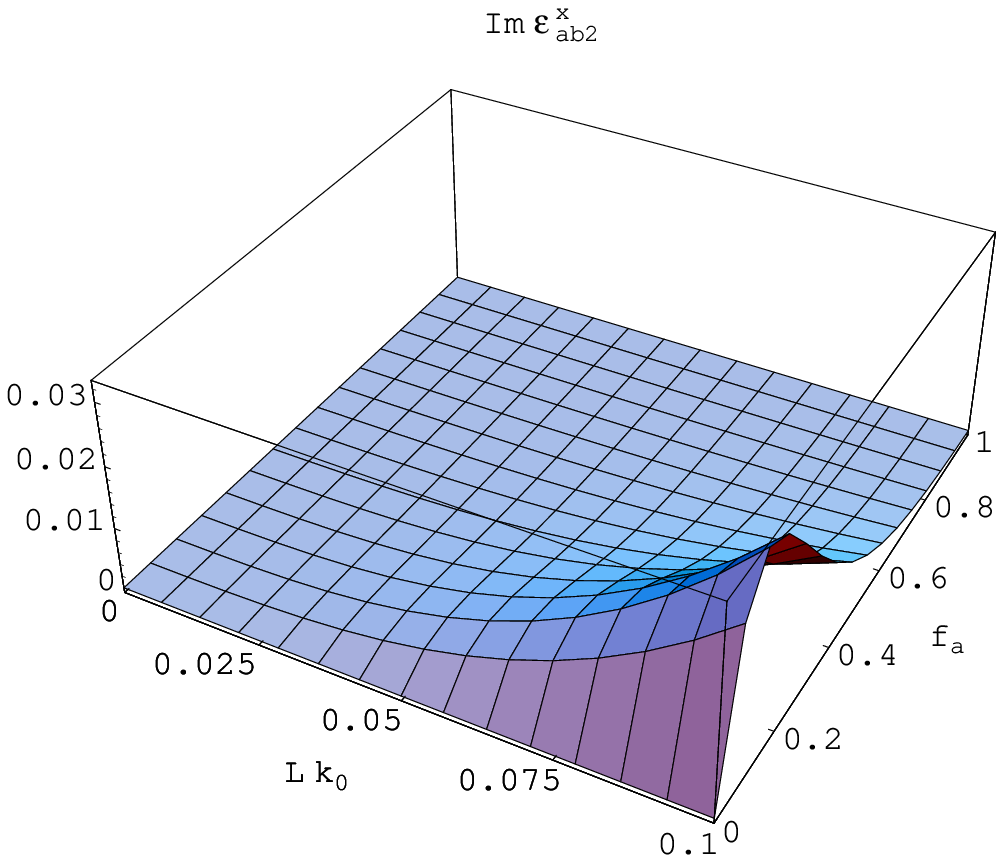,width=3.0in}
\epsfig{file=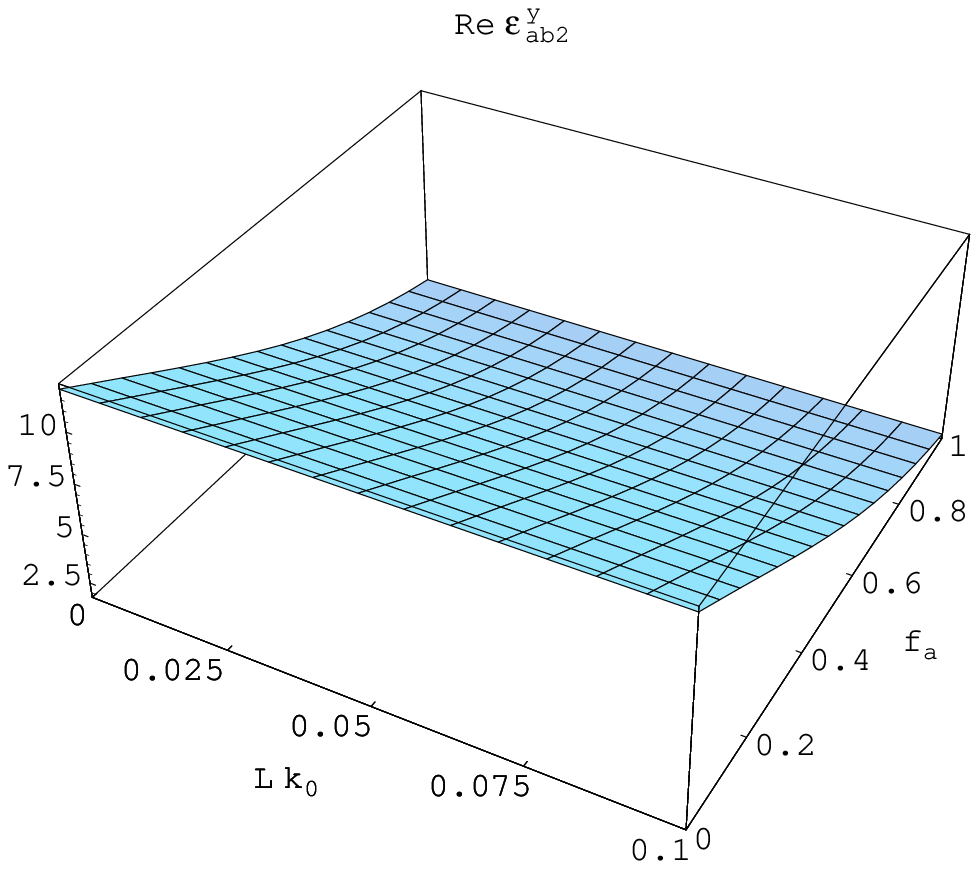,width=3.0in} \hfill
\epsfig{file=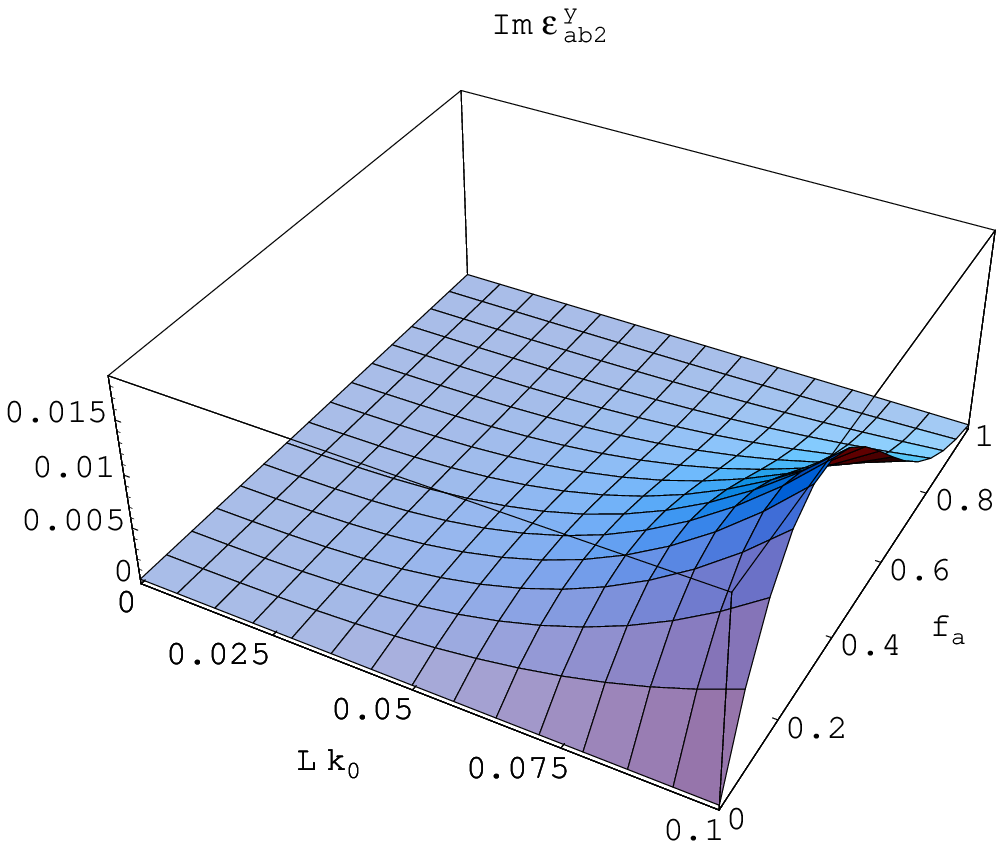,width=3.0in}
\epsfig{file=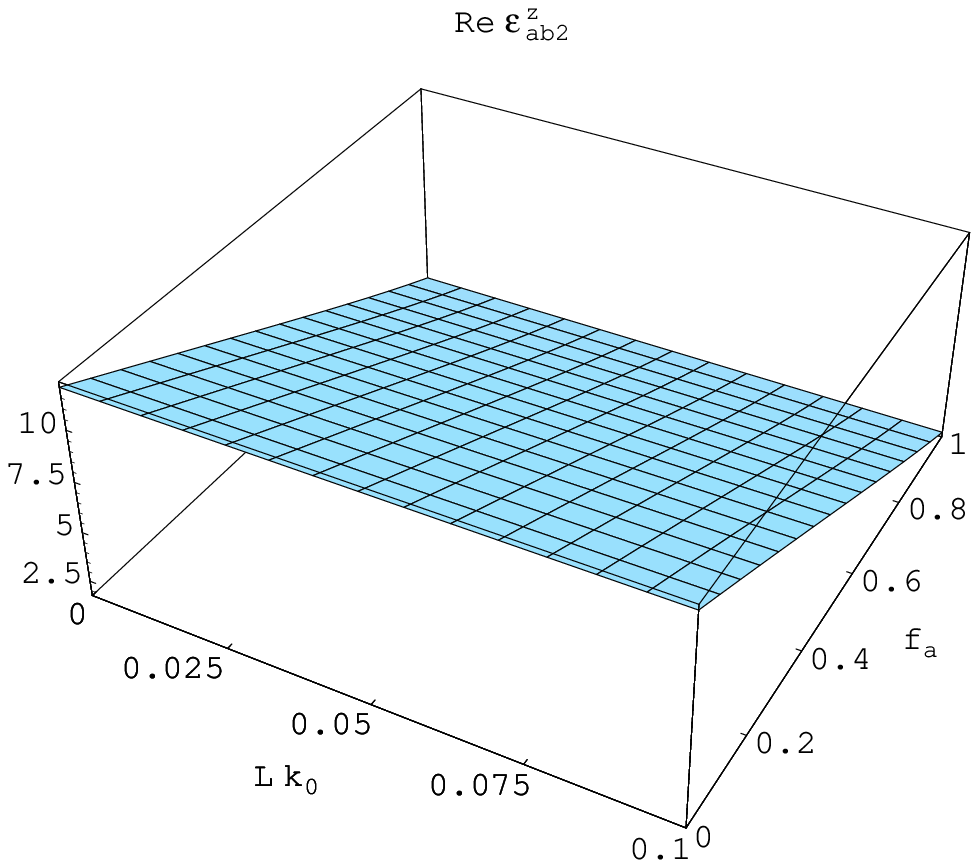,width=3.0in} \hfill
\epsfig{file=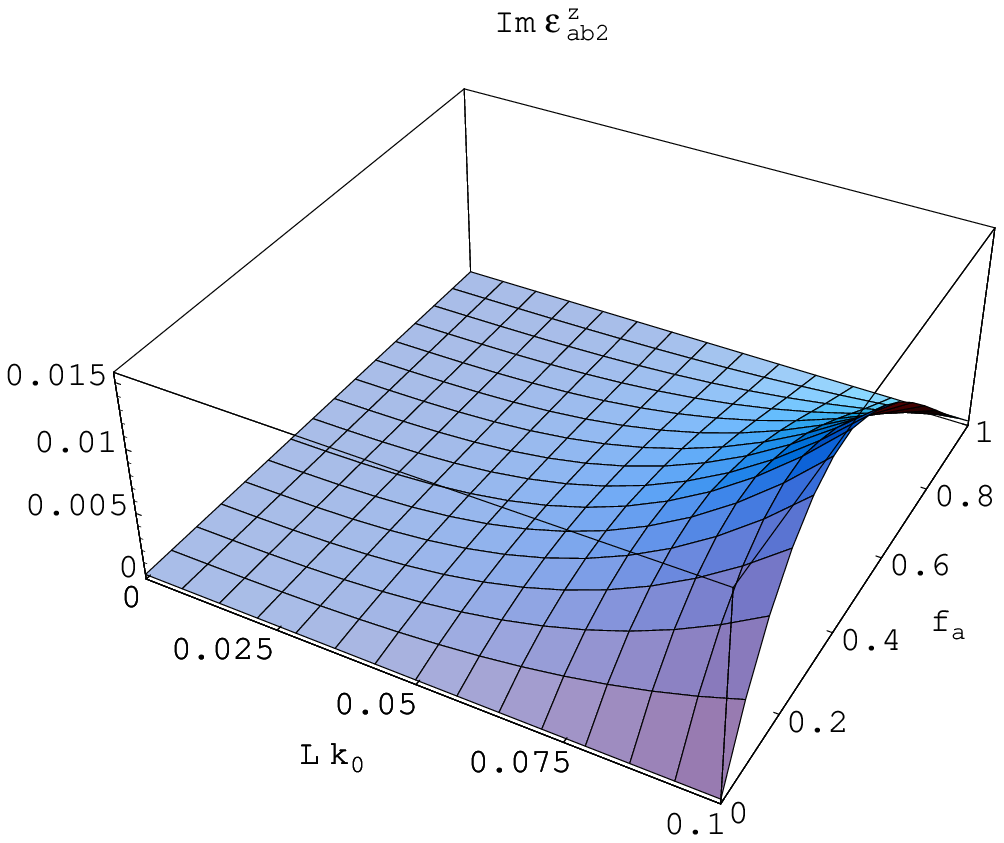,width=3.0in} \caption{ Real (left) and
imaginary (right) parts of the second order SPFT estimate of HCM
permittivity $\, \underline{\underline{ \eps}}^{[2]}_{\,ab} =
\epso \, \mbox{diag} \le \eps^x_{ab2}, \,\eps^y_{ab2},
\,\eps^z_{ab2} \ri $   plotted against volume fraction $f_a$ and
relative correlation length  $\ko L $, with inclusion size
parameter $\eta = 0$.
 Component phase
permittivities  $\eps_{a }$ and  $\eps_b $, and shape parameters
 $U_x $, $U_y $ and $U_z $,  as in Figure~1.}
\end{figure}

\newpage

\begin{figure}[!ht]
\centering \psfull \epsfig{file=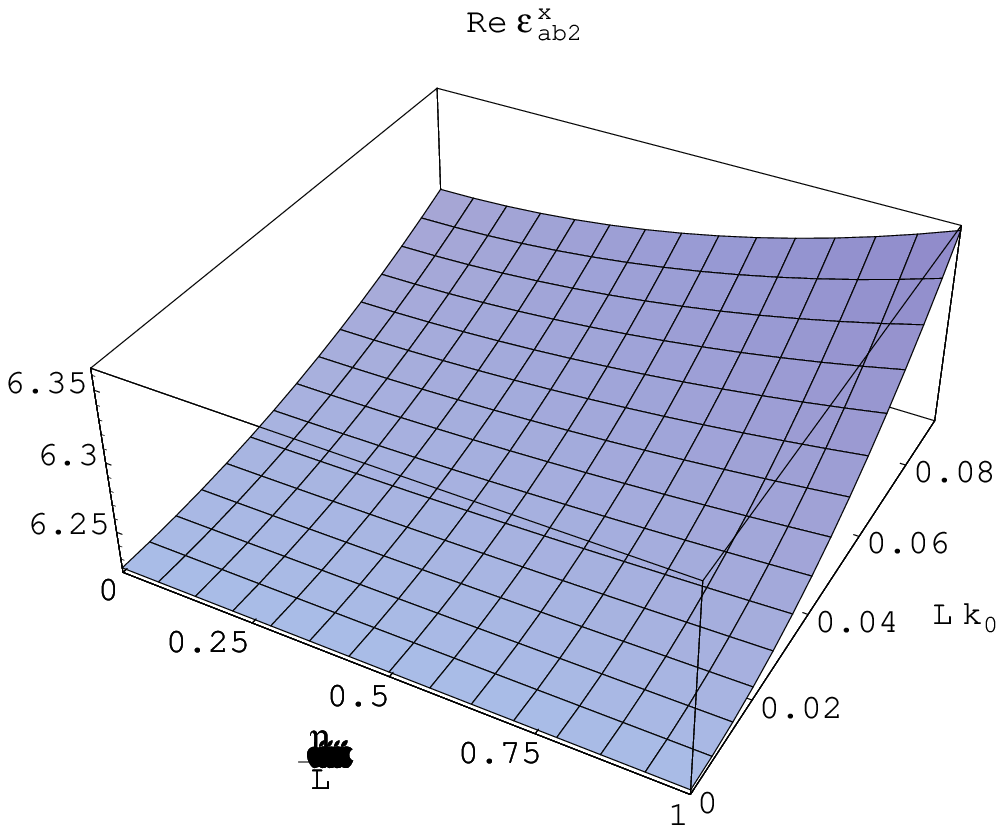,width=3.0in}
\hfill
  \epsfig{file=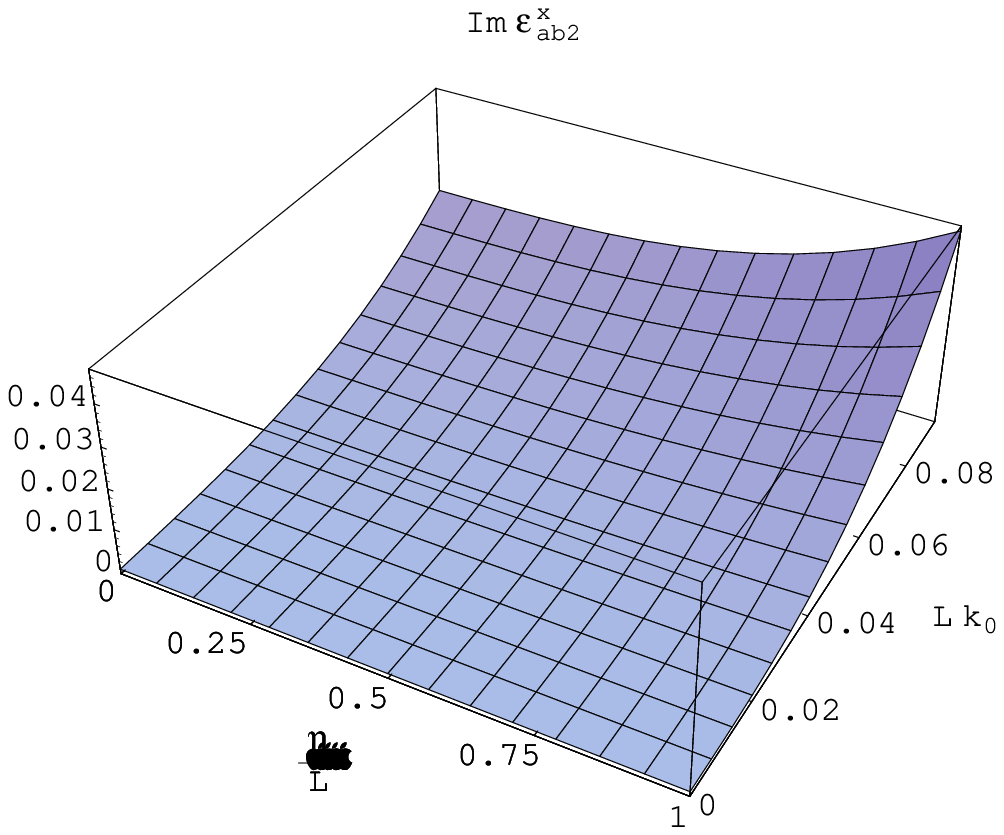,width=3.0in}
\epsfig{file=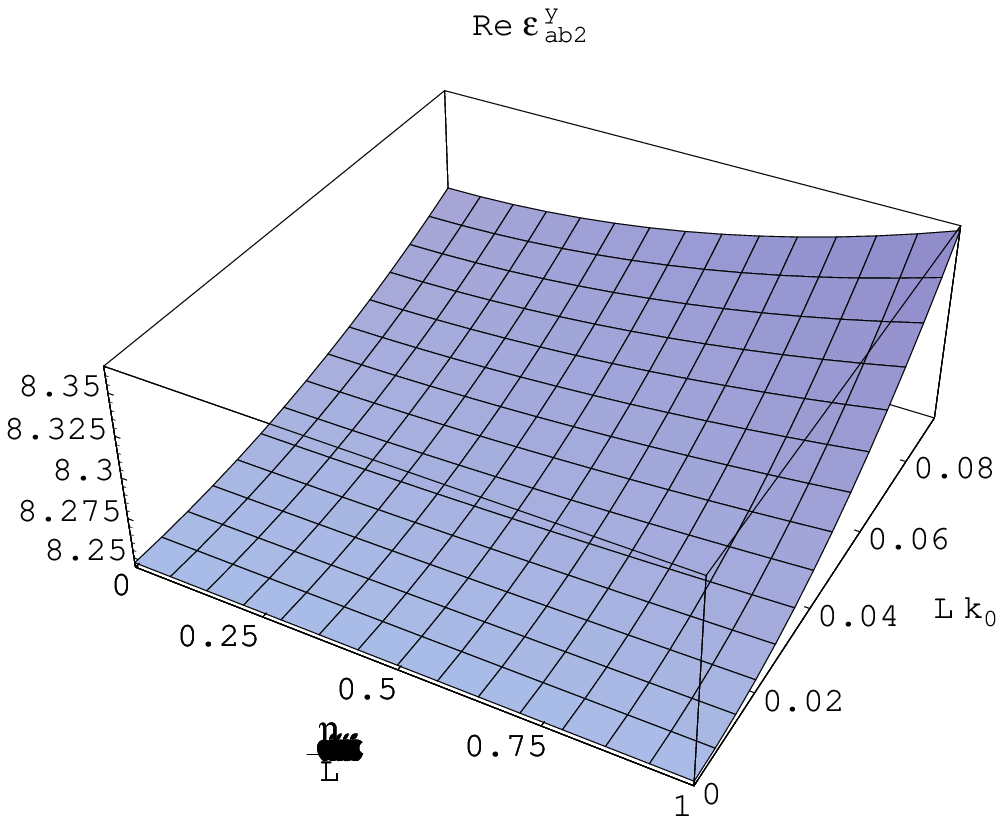,width=3.0in} \hfill
\epsfig{file=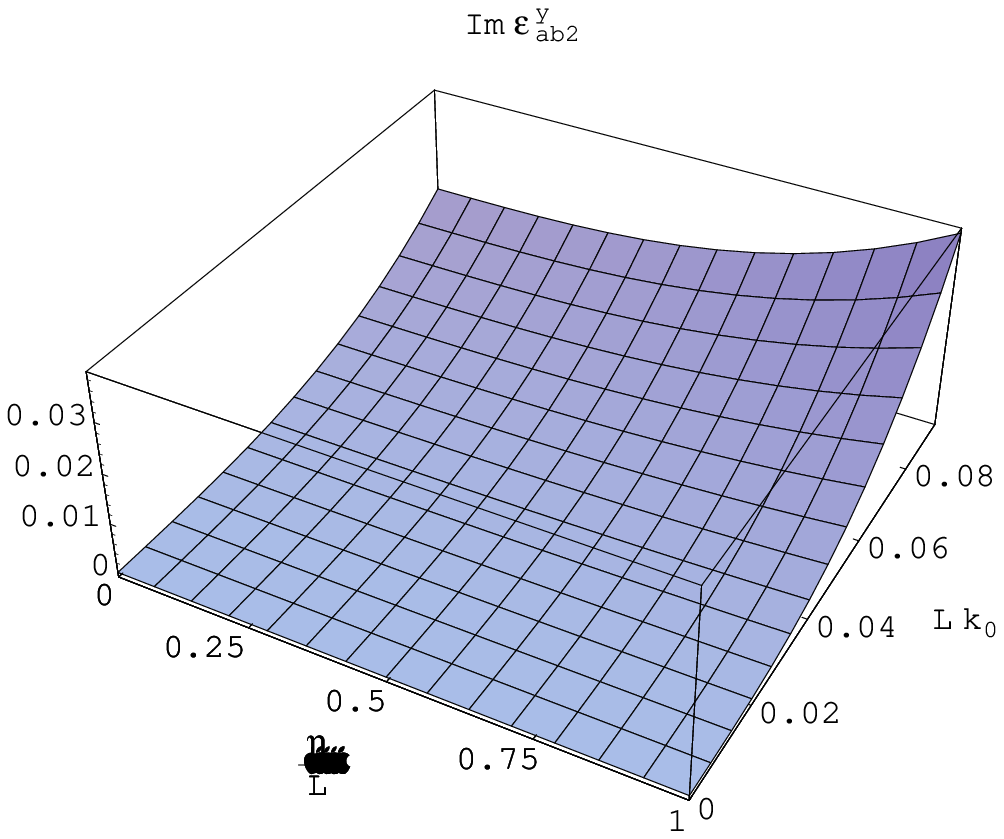,width=3.0in}
\epsfig{file=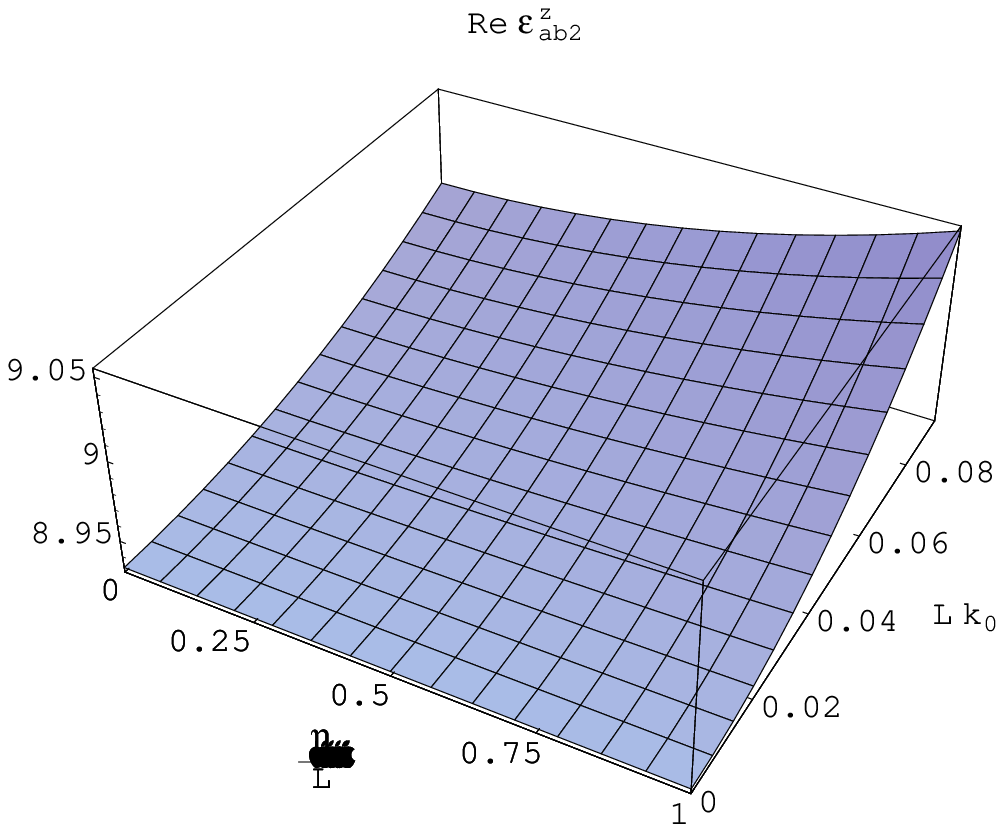,width=3.0in} \hfill
\epsfig{file=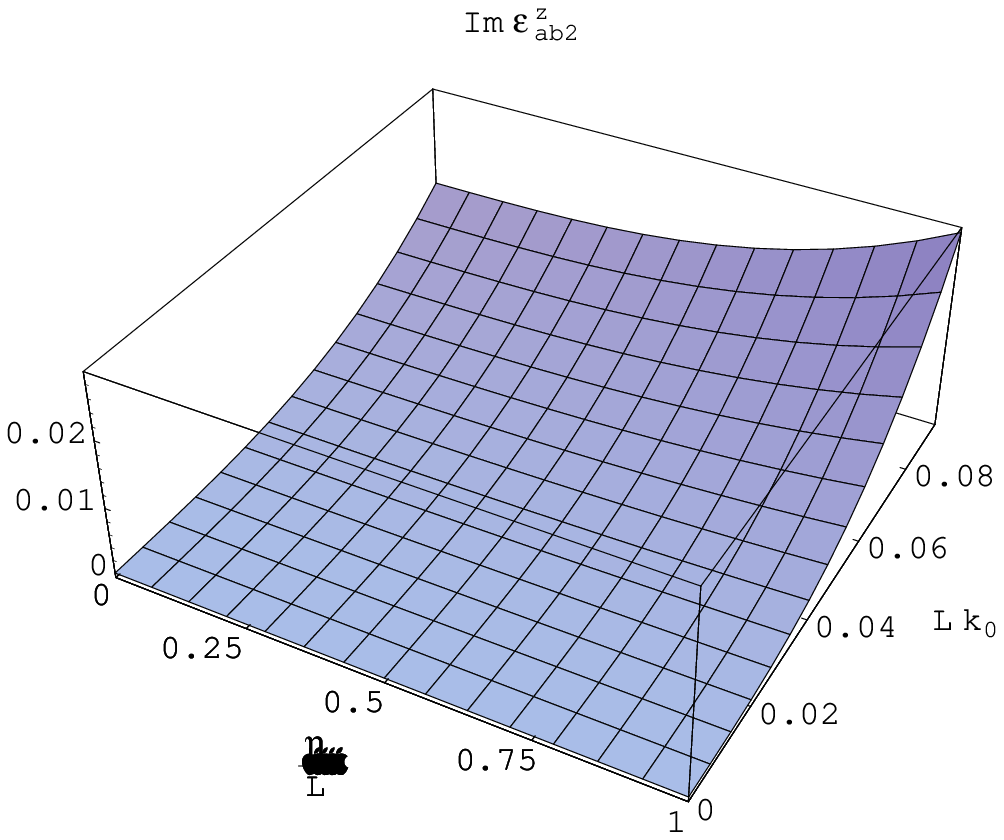,width=3.0in} \caption{ Real (left) and
imaginary (right) parts of the second order SPFT estimate of HCM
permittivity $\, \underline{\underline{ \eps}}^{[2]}_{\,ab} =
\epso \, \mbox{diag} \le \eps^x_{ab2}, \,\eps^y_{ab2},
\,\eps^z_{ab2} \ri $ plotted relative correlation length  $\ko L $
and relative inclusion size parameter  $\eta / L$, with volume
fraction $f_a = 0.3$.
 Component phase
permittivities  $\eps_{a }$ and  $\eps_b $, and shape parameters
 $U_x $, $U_y $ and $U_z $,  as in Figure~1.
}
\end{figure}

\newpage

\begin{figure}[!ht]
\centering \psfull \epsfig{file=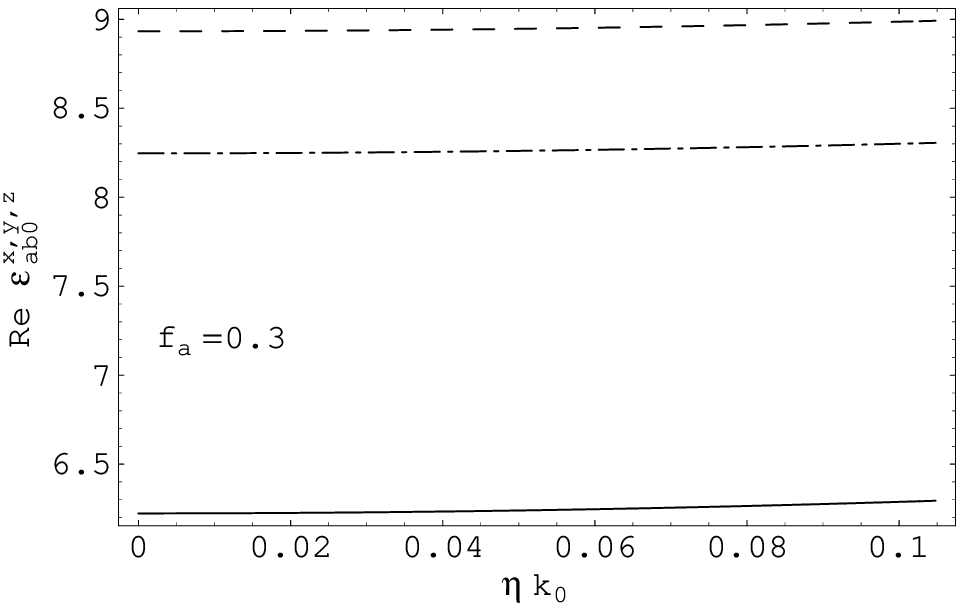,width=5.0in}
  \epsfig{file=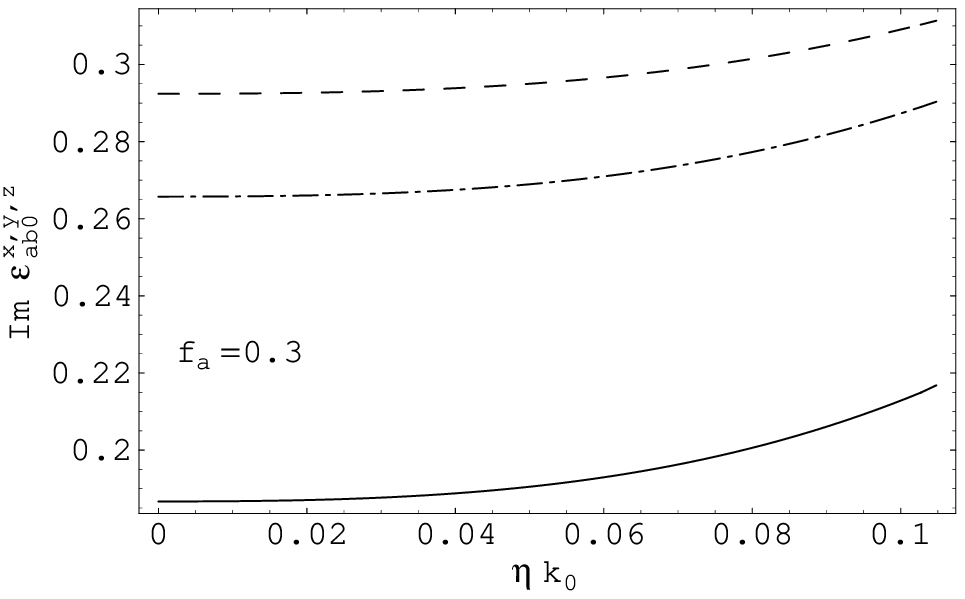,width=5.0in}
\caption{ Real (top) and imaginary (bottom) parts of the zeroth
order SPFT estimate of HCM permittivity
 $\, \underline{\underline{ \eps}}^{[0]}_{\,ab} = \epso \,
\mbox{diag} \le \eps^x_{ab0}, \,\eps^y_{ab0}, \,\eps^z_{ab0} \ri $
  plotted against
 relative inclusion size $\ko \eta $ for volume fraction $f_a =0.3$.
 Component phase
parameter values: $\eps_{a } = \le 2 + 0.05i \ri \epso$,  $\eps_b
 = \le 12 + 0.4i \ri \epso$, $U_x = 1$, $U_y = 3$ and $U_z = 15$.
Key: $ \eps^x_{ab0}$, $\eps^y_{ab0}$ and $ \eps^z_{ab0}$ are denoted
by
  solid, broken dashed and  dashed lines respectively.
}
\end{figure}

\newpage

\begin{figure}[!ht]
\centering \psfull \epsfig{file=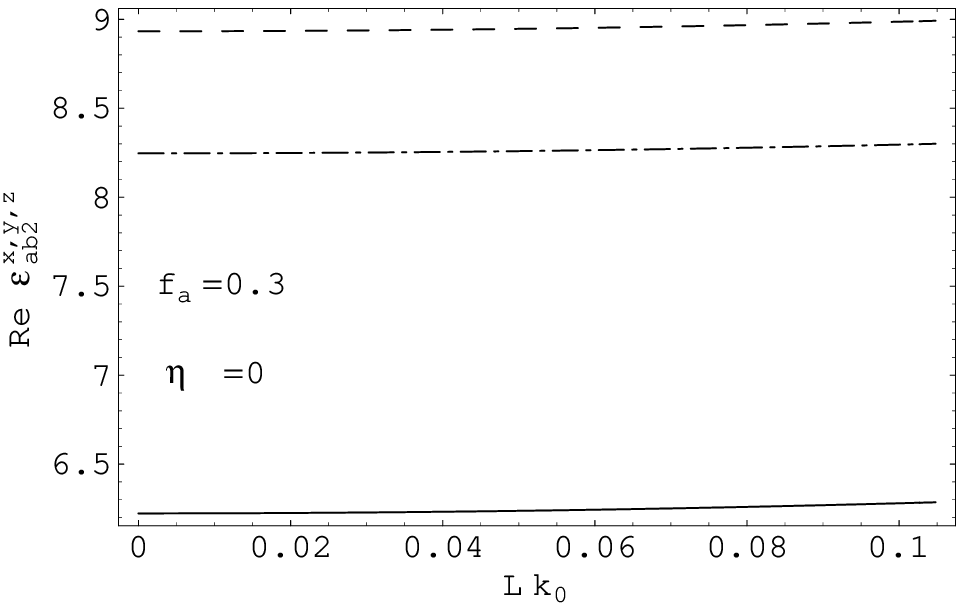,width=5.0in}
  \epsfig{file=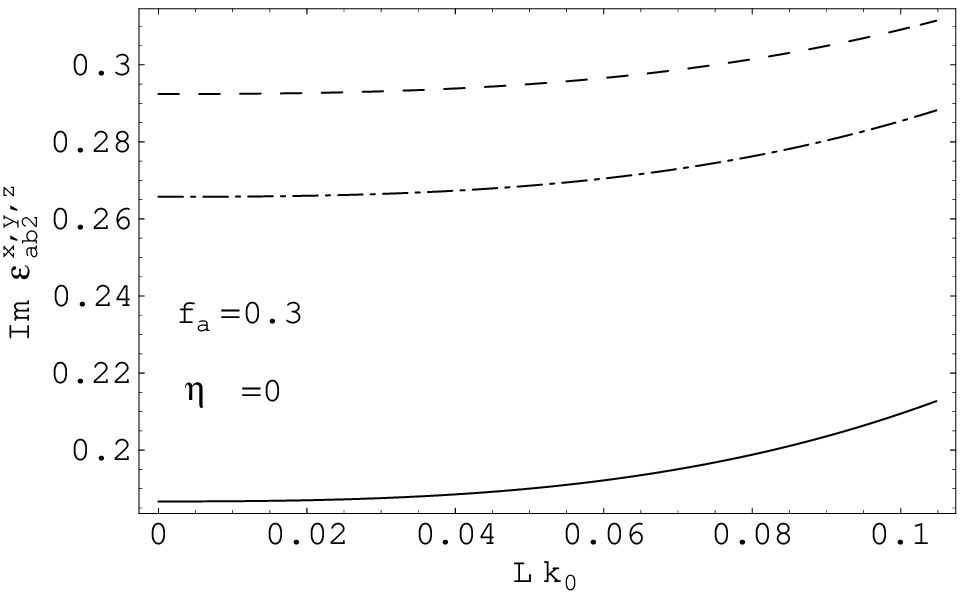,width=5.0in}
\caption{ Real (top) and imaginary (bottom) parts of the second
order SPFT estimate of HCM permittivity $\, \underline{\underline{
\eps}}^{[2]}_{\,ab} = \epso \, \mbox{diag} \le \eps^x_{ab2},
\,\eps^y_{ab2}, \,\eps^z_{ab2} \ri $   plotted against
 relative correlation length $\ko L $,
with  inclusion size parameter $\eta = 0$
and
volume fraction $f_a =0.3$.
 Component phase
permittivities  $\eps_{a }$ and  $\eps_b $, and shape parameters
 $U_x $, $U_y $ and $U_z $,  as in Figure~4.
Key: $ \eps^x_{ab2}$, $\eps^y_{ab2}$ and $ \eps^z_{ab2}$ are denoted
by
  solid, broken dashed and  dashed lines respectively.
}
\end{figure}

\newpage

\begin{figure}[!ht]
\centering \psfull \epsfig{file=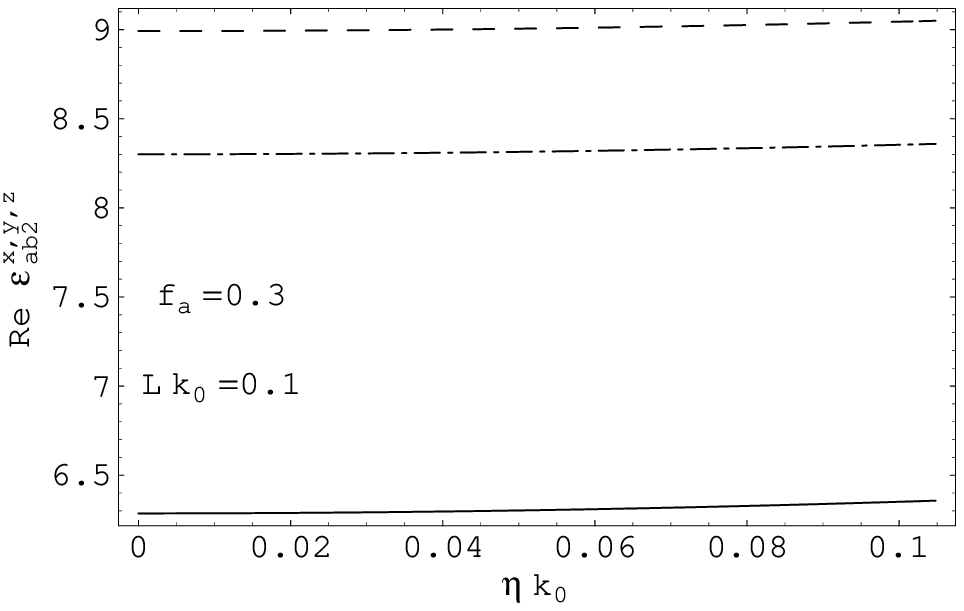,width=5.0in}
  \epsfig{file=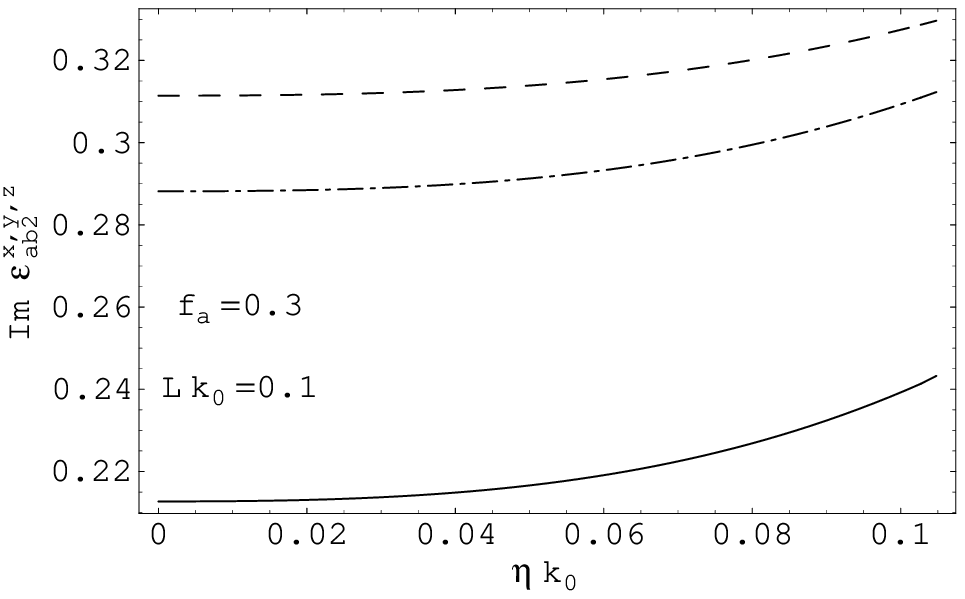,width=5.0in}
\caption{ Real (top) and imaginary (bottom) parts of the second
order SPFT estimate of HCM permittivity $\, \underline{\underline{
\eps}}^{[2]}_{\,ab} = \epso \, \mbox{diag} \le \eps^x_{ab2},
\,\eps^y_{ab2}, \,\eps^z_{ab2} \ri $   plotted relative
 inclusion size parameter $\ko \eta $, with
 relative
 correlation length $\ko L = 0.1 $ and
volume
fraction $f_a = 0.3$.
 Component phase
permittivities  $\eps_{a }$ and  $\eps_b $, and shape parameters
 $U_x $, $U_y $ and $U_z $, as in Figure~4; and key as in Figure~5.
}
\end{figure}

\end{document}